%% file: structure_theorem.tex
\documentclass[11pt,letterpaper]{article}
\pdfoutput=1
\usepackage{amsmath,amssymb,amsfonts,amsthm}
\usepackage{latexsym,nicefrac,bbm}
\usepackage{xspace}
\usepackage{fullpage}
\usepackage{color}
\usepackage{hyperref}

\input{macros}

\title{A reformulation of the Arora-Rao-Vazirani Structure Theorem}
\author{Sanjeev Arora\thanks{Department of Computer Science and Center for Computational Intractability, Princeton University. Supported by NSF Grants
CCF-0832797, 0830673, and 0528414.} \and
James Lee\thanks{Computer Science and Engineering, University of Washington.} \and
Sushant Sachdeva\thanks{Department of Computer Science and Center for Computational Intractability, Princeton University. Supported by NSF Grants
CCF-0832797, 0830673, and 0528414.}}

\begin{document}
\maketitle
\begin{abstract}
In a well-known paper\cite{ARV}, Arora, Rao and Vazirani obtained an
$O(\sqrt{\log n})$ approximation to the Balanced Separator problem and
Uniform Sparsest Cut. At the heart of their result is a geometric
statement about sets of points that satisfy triangle inequalities,
which also underlies subsequent work on approximation algorithms and
geometric embeddings.

In this note, we give an equivalent formulation of the \emph{Structure theorem} in \cite{ARV} in terms of the expansion of large sets in geometric graphs on sets of points satisfying triangle inequalities.
\end{abstract}

\section{Introduction}

\begin{Def}[Triangle Inequalities]
A set of points $V$ is said to satisfy \emph{triangle inequalities} if for every $v_i,v_j,v_k \in V$, the following inequality holds
\[\|v_i-v_j\|^2 + \|v_j-v_k\|^2 \ge \|v_i - v_k\|^2\]
\end{Def}

For a set of points $V$, we define \emph{average squared distance} to
be the expression $\av_{i,j} \|v_i-v_j\|^2$ where the expectation is taken over all pairs $i,j \in V$ . 
The following geometric theorem was shown in the well-known paper by
Arora, Rao and Vazirani\cite{ARV}. This theorem and its variants underlie subsequent work on improved approximation algorithms for several fundamental problems \cite{ACMM,Kar,CMM} and metric embeddings \cite{CGR,ALN}.

\begin{Thm}[ARV structure theorem (Theorem 1) \cite{ARV}; existence of ``well-separated sets'']
\label{ARVthm}
 For every $c > 0$ there exist $c^\prime, b > 0$ such that the following holds for all $n,d$: Given $n$ points on unit $(d-1)$-sphere, $v_1,\ldots,v_n
\in \rea^d$ that satisfy triangle inequality such that the \emph{average squared distance} is $c$, then there exist two sets $S,T \subseteq \{v_i\}_{i \in [n]}$ of size at least $c^\prime n$ such that for every $v_i \in S, v_j \in T, \|v_i - v_j\|^2 \ge \frac{b}{\sqrt{\log n}}$
\end{Thm}

There has been subsequent work on efficient algorithms for Uniform Sparsest Cut and the Balanced separator problem \cite{AHK,AK,Sherman}. These results require efficient algorithmic variants of the structure theorem and are based on the notion of \emph{expander flows}\cite{ARV}.

Our equivalent formulation concerns the expansion of some family of geometric
graphs. Now we define this family.

\begin{Def}[$G_{V,\epsilon}$]
Given a set of points $V \subseteq \rea^d$, we define $G_{V,\epsilon}$ to be the graph on the vertex set $V$ obtained by adding an edge between any two points $v_i,v_j$ such that $\|v_i - v_j\|^2 \le \epsilon$.
\end{Def}

For any $\epsilon \ge 0$, $G_{V,\epsilon}$ has a self-loop at each vertex. Thus if $\Gamma(S)$ denotes the set of neighbors of $S$, $S\subseteq \Gamma(S)$.

Our reformulation will talk about the expansion of large sets in graphs $G_{V,\epsilon}$. We will use the following definition of an \emph{expander}. This definition is not really standard but has been tailored to improve readability.

\begin{Def}
 A graph $G$ is said to be an $(\alpha,\beta)$-expander if for every set $S$ of
size $\alpha|V(G)| \le |S| \le \frac{1}{2\beta}|V(G)|$, we have $|\Gamma(S)| > \beta|S|$ (where $\Gamma(S)$ denotes the set of neighbors of $S$). 
\end{Def}

Note that the definition requires a lower bound on the size of the set $S$. Also note the strict inequality in the requirement for the size of $\Gamma(S)$. For the graphs that we care about, $S \subseteq \Gamma(S)$ and hence $\beta > 1$ for the definition to be non-trivial.

Observe that an $(\alpha,\beta)$-expander is also an $(\alpha^\prime,\beta^\prime)$-expander for all $\alpha^\prime \ge \alpha$ and $\beta^\prime \le \beta$.

In this note we prove that the following is an equivalent reformulation of the
ARV Structure theorem.

\begin{Thm}[Main]
\label{survey}
 For every $c > 0$, there exist $\gamma > 0, \frac{1}{2} > \alpha > 0$
 such that the following holds for all $n,d,\epsilon$ and $\beta > 1$:
 Given a set $V$ of $n$ points on the unit sphere that satisfy the triangle inequality condition such that their
average squared distance is at least $c$, and $G_{V,\epsilon}$ is an
$(\alpha,\beta)$-node expander, then $\epsilon \ge \frac{\gamma}{k\sqrt{\log n}}$ where $k = \ceil{\log_\beta \frac{1}{2\alpha}}$ (or equivalently $n \geq exp(\gamma^2/k^2\epsilon^2)$).
\end{Thm}

\section{Proof of the main theorem}
\begin{proof}
($\ref{survey} \implies \ref{ARVthm}$) Given a set $V$ of $n$ points on the unit sphere that satisfies the conditions of theorem \ref{ARVthm}, we construct the graph $G_{V,\epsilon}$ for $\epsilon = \frac{\gamma}{3\sqrt{\log n}}$. Now, using theorem $\ref{survey}$ with the same $c$ and $\beta^2 = \frac{1}{2\alpha}$, there exists a non-trivial $\alpha$ such that $G_{V,\epsilon}$ is not an $\left(\alpha,\sqrt{\frac{1}{2\alpha}}\right)$-expander (since $k=2$ and $\epsilon < \frac{\gamma}{2\sqrt{\log n}}$). 

Thus, there exists a set $S$ such that $\alpha n \le |S| \le \frac{1}{2\beta}n = \sqrt{\frac{\alpha}{2}}n$ such that $\Gamma(S) \le \beta|S| \le \frac{1}{2}n$. This means that there is a set $T = V \backslash \Gamma(S)$ of size at least $\frac{1}{2}n$ such that there are no edges between $S$ and $T$. 

Thus $S$ and $T$ are sets of size at least $c^\prime n$ (for $c^\prime = \alpha$) such that there is no edge between them in $G_{V,\epsilon}$. Hence they are $\frac{b}{\sqrt{\log n}}$-separated for $b=\frac{\gamma}{3}$ (by the definition of $G_{V,\epsilon}$).

\ \\

 ($\ref{ARVthm} \implies \ref{survey}$) Given $n$ points on the unit sphere that
satisfy the conditions of theorem \ref{survey}, we use theorem \ref{ARVthm} with the same $c$ to get two sets $S,T$ of size at least $c^\prime n$ such that they are $\frac{b}{\sqrt{\log n}}$ separated. Without loss of generality, we shall assume that $|S| \le |T|$.

Assume there is some $\epsilon > 0$ such that $G_{V,\epsilon}$ is a $(c^\prime,\beta)$-expander. This implies that $|\Gamma^t(S)| > \min\{\beta^t c^\prime,\frac{1}{2\beta}n\}$ for all $t \ge 0$. Thus, denoting $k = \ceil{\log_\beta \frac{1}{2c^\prime}}$, we get $|\Gamma^{k-1}(S)| \ge \frac{1}{2\beta}n$. Pick a subset $S^\prime \subset |\Gamma^{k-1}(S)|$ such that $|S| = \frac{1}{2\beta}n$. Now $\Gamma^k(S) \supseteq \Gamma(S^\prime)$ and hence $|\Gamma^k(S)| > \frac{1}{2}n$. 

Similarly $|\Gamma^k(T)| > \frac{1}{2}n$. Since $|\Gamma^k(S)|,|\Gamma^k(T)| > \frac{1}{2}n$, we must have $|\Gamma^k(S) \cap \Gamma^k(T)| > 0$ giving a path of length $2k$ between a vertex in $S$ and a vertex in $T$. Since $S$ and $T$ are $\frac{b}{\sqrt{\log n}}$ separated, using the fact that triangle inequalities imply that $\|\cdot\|^2$ is a metric on $V$, we get that $2k\epsilon \ge \frac{b}{\sqrt{\log n}}$. This gives us the statement of the theorem with $\alpha = c^\prime,\gamma = \frac{b}{2}$.
\end{proof}

The notion of expansion used in the above theorem is tight in the sense that if we wish to consider a $(\cdot,\beta)$-expander, we require that sets of size $\frac{1}{2\beta}n$ should expand. Consider the following example: Place half the vertices at one point on the unit sphere and
the remaining half on the antipodal point on the sphere. This set of vertices
satisfies the triangle inequality and the average squared distance between any two
vertices is a constant. Also, for any $\epsilon \ge 0$, the graph
$G_{V,\epsilon}$ consists of two cliques, each consisting of half the vertices.
Thus, we cannot lower bound $\epsilon$ if being an $(\alpha,\beta)$-expander does not rule out the possibility of having two equal sized disjoint
cliques. So we require the expansion to guarantee that $|\Gamma^k(S)| > \frac{1}{2}n$ for some $k$.

\section{Additional comments}
The proof of the structure theorem uses triangle inequality only to deduce that if for every $i, \|x_i - x_{i+1}\|^2 \le \Delta$, then $\|x_1 - x_k\|^2 \le k\Delta$. This simple fact is used for $k=O(\sqrt{\log n})$. In fact, a slightly stronger theorem that does not explicitly require triangle inequalities can be deduced from their proof. 

Letting $B_2^d$ denote the $d$-dimensional ball, we have the following theorem:
\begin{Thm}
For every $c > 0$, there exist $\delta,\gamma > 0, \frac{1}{2} > \alpha > 0$
 such that the following holds for all $n,d,\epsilon$ and $\beta > 1$:
Given an $n$-vertex graph $G = (V,E)$ that is an $(\alpha,\beta)$-expander and assume that $f : V \to B_2^d$ satisfies that the average squared distance is at least $c$, then there are $u,v \in V$ such that 
\[d_G(u,v) \le \gamma\frac{\sqrt{\log n}}{\log \beta} \text{ and } \|f(u)-f(v)\|_2 \ge \delta\]
\end{Thm}

We do not know of a simple way to derive this from the structure theorem as stated but it is easy to deduce the above theorem from their proof by keeping in mind that we want to replace the triangle inequality requirement with a requirement on all paths of length $O(\sqrt{\log n})$ in the graph. 

This is similar to a theorem used to prove the existance of expander flows (Lemma 28,\cite{ARV}). A similar theorem has been stated in $\cite{NRS}$ (Proposition 3.11) and proved using essentially the same proof as of the structure theorem from $\cite{ARV}$.

\bibliographystyle{plain}
\bibliography{structure_theorem}
\end{document}

%% file: macros.tex
\newcommand\rea{\mathbb R}

\newcommand{\marginlabel}[1]%
{\mbox{}\marginpar{\it{\raggedleft\hspace{0pt}#1}}}

\newcommand{\ceil}[1]{\left\lceil\, {#1}\,\right\rceil}

\definecolor{Mygray}{gray}{0.8}

 \ifcsname ifcommentflag\endcsname\else
  \expandafter\let\csname ifcommentflag\expandafter\endcsname
                  \csname iffalse\endcsname
\fi

\ifcommentflag

\else

\fi

\ifcommentflag
\newcommand{\Authornote}[2]{{\sf\small\color{red}{[#1: #2]}}}
\newcommand{\Authoredit}[2]{{\sf\small\color{red}{[#1]}\color{blue}{#2}}}
\newcommand{\Authorcomment}[2]{{\sf \small\color{red}{[#1: #2]}}}
\newcommand{\Authorfnote}[2]{\footnote{\color{red}{#1: #2}}}
\newcommand{\Authorfixme}[1]{\Authornote{#1}{\textbf{??}}}
\newcommand{\Authormarginmark}[1]{\marginpar{\textcolor{red}{\fbox{
#1:!}}}}
\else
\newcommand{\Authornote}[2]{}
\newcommand{\Authoredit}[2]{}
\newcommand{\Authorcomment}[2]{}
\newcommand{\Authorfnote}[2]{}
\newcommand{\Authorfixme}[1]{}
\newcommand{\Authormarginmark}[1]{}
\fi

\def\implies{\Rightarrow}

\newcommand\av{\mathop{\mbox{\bf E}}}

\newlength{\pgmtab}  
\newtheorem{Thm}{Theorem}[section]

\newtheorem{Def}[Thm]{Definition}

 {
	\begin{enumerate}}{\end{enumerate}}

\newcounter{lecnum}

\newlength{\tpush}
